\documentclass[printer]{aa}
\usepackage[varg]{txfonts}

\usepackage{graphicx}
\usepackage{natbib}
\usepackage{color}
\usepackage{subfigure}
\usepackage[switch]{lineno} 
\usepackage{xspace}
\usepackage{microtype}
\usepackage{amssymb}
\def\link_col{blue}
\usepackage{xspace}
\usepackage{url}
\usepackage{wasysym}

\def\fermi{{\it Fermi}-LAT\xspace}

\def\gray{$\gamma$-ray\xspace}

\begin{document}

\title{On the  GeV excess  in the diffuse \gray emission towards the Galactic centre  }
\titlerunning{On the  GeV excess  in the diffuse \gray emission}
\author{Rui-zhi Yang\inst{1}
\and Felix Aharonian\inst{1, 2, 3}}
\institute{Max-Planck-Institut f{\"u}r Kernphysik, P.O. Box 103980, 69029 Heidelberg, Germany.
\and Dublin Institute for Advanced Studies, 31 Fitzwilliam Place, Dublin 2, Ireland.
\and MEPHI, Kashirskoe shosse 31, 115409 Moscow, Russia
}%

\date{Received:  / Accepted: } 
\abstract {}
{
The \fermi  $\gamma$-ray data have been used  to study 
the morphological and spectral features of the so-called  GeV excess  - a 
diffuse radiation component  recently  discovered   towards the Galactic centre.}
{
We used the likelihood method to analyze \fermi data. Our study does confirm  the existence of such an 
extra  component in the  diffuse  \gray  emission at GeV energies.  Based on a detailed  
morphological analysis,  a spatial template that fits the data best  was generated and adopted.
}
{
Using this template,  the energy  distribution of  $\gamma$-rays was derived in the  0.3-30 GeV  energy interval.  The spectrum  appeared to have less distinct  (`bump'-like)  structure 
than previous reported.
We argue that the morphology of this  radiation component has a bipolar rather than a spherically symmetric structure as has been assumed a priori in previous studies.  
}
{This  finding excludes the associations of the 
GeV excess with Dark Matter.   We briefly discuss the radiation mechanisms and 
possible source populations  that could be responsible for this  new component of diffuse gamma radiation.
}


\maketitle

\section{Introduction}
The analysis of the \fermi  $\gamma$-observations by several independent groups \citep{goodenough09, ak12, daylan13, macias14, weniger14, zhou15, huang15}  has revealed  a significant  excess in the diffuse gamma radiation towards the Galactic centre (GC) compared to the one predicted by standard empirical Galactic background models.   Since the excess  has a maximum  in the spectral energy distribution (SED) of radiation  at energies of several GeV,  it was dubbed  the GeV excess.   Furthermore,  this spectral structure has been interpreted  as a hint of dark matter in the GC region \citep{goodenough09, daylan13, weniger14}.   However, this  is not a unique interpretation, given  the alternative  explanations  related, for example, to the radiation of millisecond pulsars in the bulge  \citep{yuan14, bartels15, mirabal13} or to the specific propagation effects of cosmic rays \citep{carlson14, macias14_2, cholis15}.

In this paper we focus on the inner $10^{\circ} \times 10^{\circ}$ region of the GC  and use the new 3FGL source catalogue \citep{3fgl} for a detailed analysis of the \fermi data. We do confirm the existence of an extra \gray component  revealed by previous studies.  We  explored  different residual maps  and  generated  
spatial templates  that  fit the data substantially better than the NFW profile. Remarkably, these 
templates  result in a bipolar morphology  of radiation that can hardly be explained by dark matter.  Moreover,
the  derived  energy spectrum  of `excess' $\gamma$ rays can be  interpreted  within conventional (astrophysical) models without a need to invoke dark matter. 

The paper is structured as follows.  In Section~\ref{sec:data}, we present the results of analysis of  \fermi observations and discuss the observational uncertainties and systematics of the used method. In Section~\ref{sec:discuss} we discuss  the possible origin of the extra  \gray component. The results and main conclusions are summarized in Section~\ref{sec:conc}.

\section{Data analysis}\label{sec:data}
We selected observations with \fermi  of the regions towards the GC over the period that covers 5.5 years of exposure time (MET 239557417 -- MET 415533077).
For the data reduction, we used the standard \fermi  analysis software package \emph{v9r33p0}\footnote{\url{http://fermi.gsfc.nasa.gov/ssc}}. 
In the analysis we  selected  all  $\gamma$-ray events  with energies exceeding 100~MeV. 
The region of interest (ROI) was defined to be a $10^ \circ \times 10^ \circ$ square centred on the position of GC. 
To reduce the effect of the background related to the Earth albedo, we excluded the time intervals when the Earth was in the field of view
from the analysis (more specifically,  when the centre of the field of view was  $52^ \circ$ above the zenith), as well as the time intervals when parts of the ROI were observed at zenith angles $> 100^ \circ$. 

The spectral analysis was performed based on the P7rep\_v15 version of the post-launch instrument response functions (IRFs). Both the front and back converted photons were selected.
Since  the Galactic diffuse model provided by the Fermi collaboration\footnote{gll\_iem\_v05\_rev1.fit, available at \url{http://fermi.gsfc.nasa.gov/ssc/data/access/lat/BackgroundModels.html}} already contains some anomalous excess in the GC region, we did not use it to account for the Galactic diffuse background. Instead, we used an old version of the Galactic diffuse background gal\_2yearp7v6\_v0.fit as suggested by \citet{daylan13}.
Alternatively,  we also used a diffuse background model based on the Planck opacity map \citep{planck}.  It is  assumed that the dust opacity map exactly traces the column density and a constant cosmic ray (CR)  distribution in ROI. In this regard we note that the Galprop code  \citep{galprop} predicts a  20 percent or less gradient of the CR  distribution in this compact region.  Then  the diffuse \gray emission produced by neutral pion decay and the electron bremsstrahlung becomes proportional to the gas column density and the dust opacity. We also included the inverse Compton (IC) component of the background using the Galprop code\footnote{\url{http://galprop.stanford.edu/webrun/}}, for the CR electrons distributions and the interstellar radiation fields (ISRF). 
Finally, the isotropic templates (iso\_source\_v05.txt) related to the CR contamination and the extragalactic \gray background were also included in the analysis.

To  take point-like sources into account that appear within ROI, we used the 3FGL source 
catalogue  \citep{3fgl}.  In the likelihood analysis,   the flux of point sources inside the ROI and all 
diffuse spatial templates are left as free parameters.  
\subsection{Morphological analysis}
For morphological studies, we use  only high energy (above $1$ GeV) events  as a first step.  We fit the data with the catalogue sources,  applying both the standard diffuse background model (gal\_2yearp7v6\_v0.fit + iso\_source\_v05.txt ) and the new diffuse background model (dust opacity map + galprop IC map +iso\_source\_v05.txt) with or without additional NFW template (see e.g. \citet{daylan13}). The fitting results are summarized  in Table. 1. The TS value for the extended structure is defined as $TS=-2Log(L_{NFW}/L_{null}$, where $L_{NFW} $ and $L_{null} $ are the likelihood function values for the model with and without the NFW template. 
The results show that the  additional NFW template indeed significantly improves the fitting procedure for  both diffuse background models.  Thus, our results confirm the conclusions of 
previous studies (e.g. \citet{daylan13, macias14}). This is not a surprise given that the basic assumptions
in ours and the previous  studies are  very similar,  except for including the 3FGL source catalogue in our study.
However, this cannot be used as an argument in favour of a dark matter origin of the `excess' radiation 

However, the improvement in the signal after applying the NFW template cannot be interpreted  as a 
support of the dark matter origin of the `excess' radiation.  In fact, it is not obvious that the GeV excess has the same spherical symmetric morphology as the {\it \emph{a priori}} chosen NFW template. 

\begin{figure*}
\centering
\includegraphics[width=0.4\linewidth]{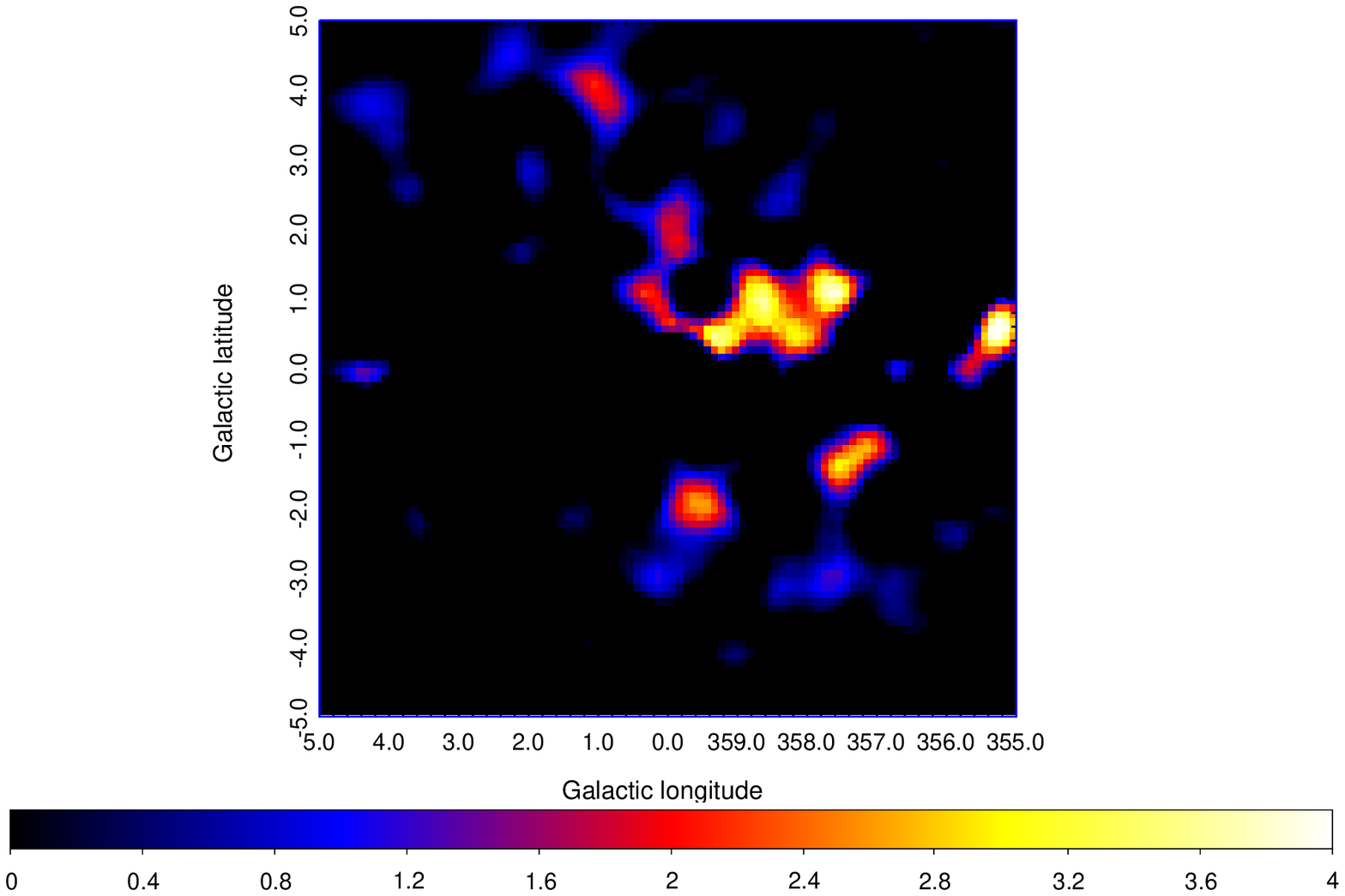}\includegraphics[width=0.4\linewidth]{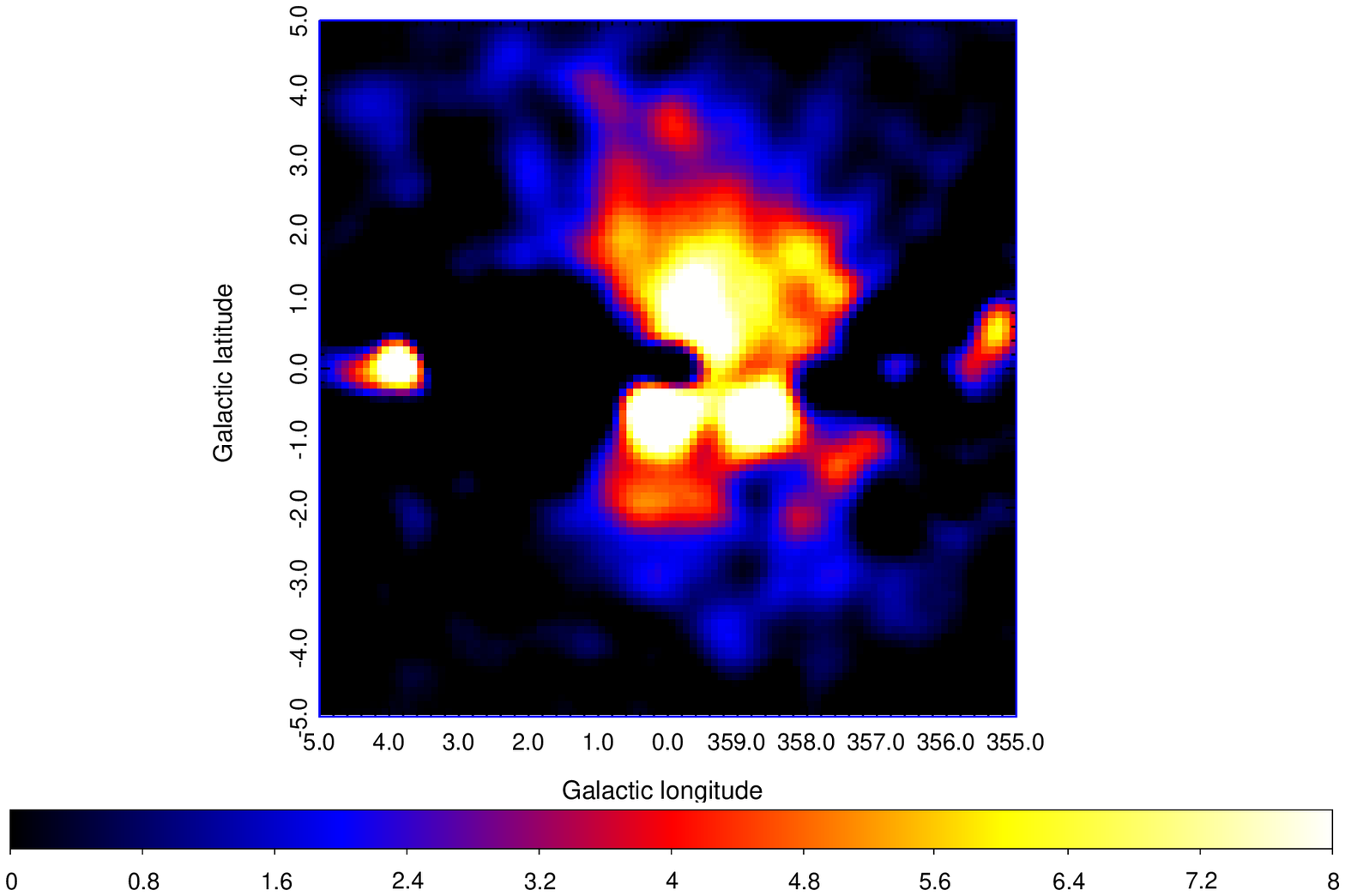}

\caption{Residual map  for  the `standard'  diffuse background model  with (left panel) and without (right panel) subtraction of the contribution from   the weak unassociated 3FGL catalogue  sources.  }
\label{fig:RES}
\end{figure*}

To pursue this issue, in Fig.\ref{fig:RES} we show the residual maps for fittings with the `standard' diffuse background model without the NFW template .  Indeed,  the residual maps reveal  significant excess to the 
northern and southern  regions,  while  there is no excess   inside the plane. This motivates us to investigate the impact of the `bipolar' morphology on the the excess radiation. Furthermore, the 3FGL catalogue includes many unassociated point sources near the GC.  To investigate alternative possibilities, such as assuming that these unassociated sources are a part of some extended structures, we omitted them when producing the residual map. The result is shown in Fig.\ref{fig:RES}. To test this hypothesis further, we derived  spatial templates based on the residual maps without the unassociated sources and used it in the likelihood fitting instead of the NFW template. The relevant estimates of the statistical significance are presented in Table.1. %

 The removal of the contamination caused by  the diffuse emission from the Galactic plane is an important part of the procedure of  extraction of the new component of radiations. The current limited understanding of both the gas and CR distributions  does not allow us  to exclude the  possibility that the bipolar structure is a result 
 of subtraction of an overestimated contribution from the Galactic plane.  On the other hand, 
it is not easy to produce a spherical symmetric residual from the data. 
This means that it would require a quite specific template, which significantly deviates from any conventional template used in the literature.  This is also confirmed by the likelihood fitting. For both diffuse background models, the bipolar residual templates reveal a significantly larger likelihood ratio compared to the spherical templates: 882 ($29~\sigma$) versus 544 ($23~\sigma$) in the standard model and 1110 ($33~\sigma$) versus 602 ($24~\sigma$) in the new diffuse model (see Table.1). Furthermore, \citet{fermigc} describe the latest Fermi official results for this region, which also show biconical residuals in the same region by using a  carefully chosen Galactic diffuse background model.  Thus we may conclude that the residual template with a bipolar structure correctly represents the spatial morphology of the GeV excess. In next section we use the standard diffuse model and the residual template as our fiducial model to derive the spectral information about the new component of the \gray emission. 

 \begin{table*}[htbp]
\caption{Fitting results for different models} \label{tab:1} \centering
\begin{tabular}{llll}
\hline
Model &\vline ~-log(likelihood)&\vline ~TS for the extended structure\\
\hline
Standard (gal\_2yearp7v6\_v0.fit + iso\_source\_v05.txt)   &\vline ~-186059 &\vline\\

\hline
Standard + NFW   &\vline  ~-186331 &\vline~544\\
\hline
Standard + residual template  &\vline  ~-186470 &\vline~822\\
\hline
New diffuse model (dust opacity map + galprop IC map +iso\_source\_v05.txt)  &\vline ~ -185423 &\vline\\
\hline
New diffuse + NFW    &\vline  ~-185724 &\vline~602\\
\hline
New diffuse +  residual template    &\vline  ~-185978 &\vline~1110\\
\hline
\end{tabular}
\end{table*}

\subsection{Spectral analysis}

\begin{figure*}
\centering
\includegraphics[width=0.4\linewidth]{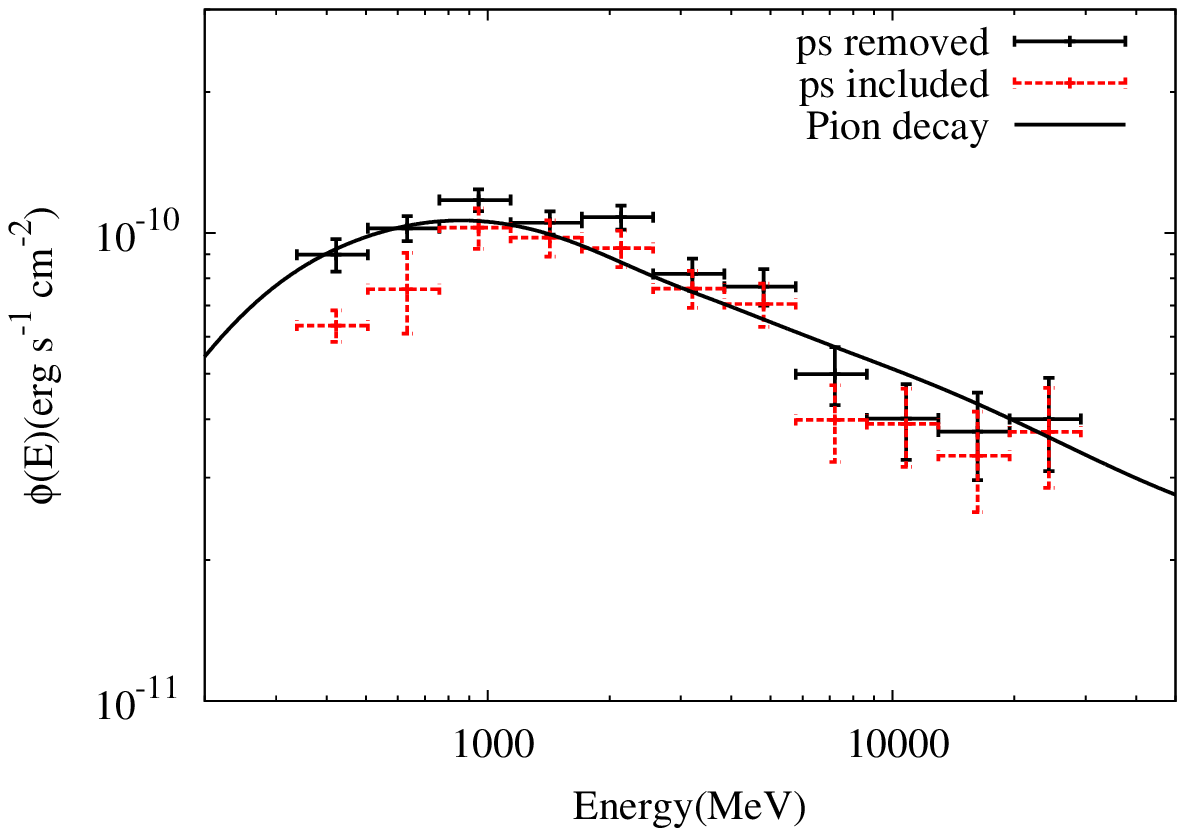}\includegraphics[width=0.4\linewidth]{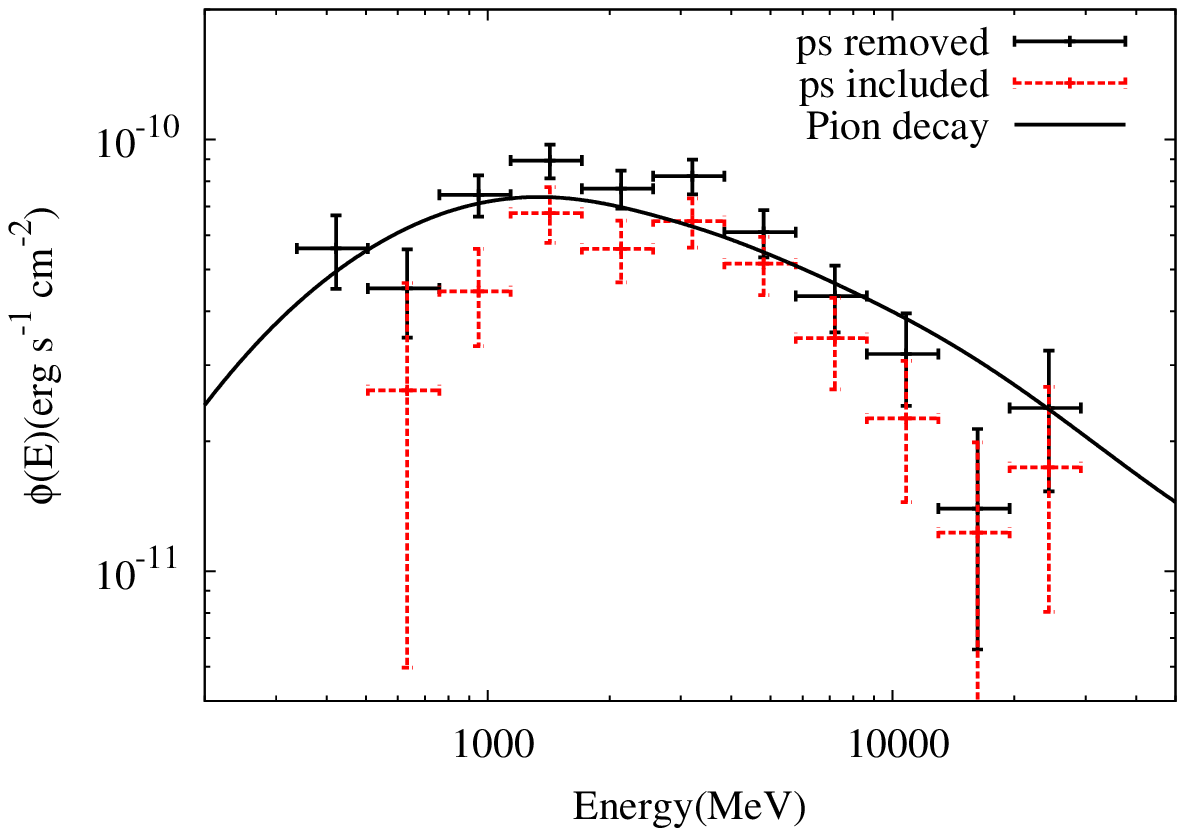}

\caption{Spectral energy distributions of {\gray}s  obtained  for  the northern (left panel) and southern (right panel) extended residual templates.  
The points  with error bars are derived from the fitting  procedure with (black symbols) and without (red symbols)   subtracting the  contribution from the  unassociated  3FGL catalogue sources. The curves represent the best spectral fits for the hadronic    ($\pi^0$ - decay) channel   (for details see the text). 
}
\label{fig:SED}
\end{figure*}

To find the spectral energy distribution (SED) of the excess, we divided the entire  energy range $300~ {\rm MeV} - 30~{\rm GeV}$ into eleven logarithmically equal energy intervals  and applied the \emph{gtlike} tool to each of these  intervals.
The results are shown in Figure~\ref{fig:SED}.
All  data points have test statistic (TS) values greater than 4, which corresponds to a statistical significance of more than $2\sigma$. 
The spectra of both the northern and southern regions show maximum of SED  at about 2 GeV,  but above 2 GeV the spectrum of the northern  part is slightly harder than that of the southern part ( $2.4 \pm 0.1$ versus $2.7 \pm 0.1$ ).  To test the possible difference between the energy spectral  of these two parts further, we tried to fit both SED above 2 GeV with the same power law index but with different  absolute fluxes. This procedure 
gives $\chi^2 / d.o.f = 14/9$, which corresponds to a $p$ value of 0.12.  Thus the hypothesis that the two parts have the same spectral shape is rejected at the level of  $1.5~\sigma$. The different spectra of the northern and southern  parts would be an independent argument against the interpretation of the gamma-ray fluxes by dark matter, but for the available data set,  the evidence  is only marginal.  
Finally, it should be noted 
that  at low energies, the spectra of  the northern and southern 
components are softer than the spectrum derived when the NFW template is used. 
In Figure~ \ref{fig:SED} we show the results from the likelihood fitting with and  without unassociated point sources. The difference between the  two cases  does not exceed 10\% in any energy range for the 
northern  part. However, for the first energy bin of the spectrum of the southern part, the difference is significantly larger. As a result, the photon index below the peak in the spectrum of the southern  part ranges between 0.5 and 2. 

\begin{figure*}
\centering
\includegraphics[width=0.4\linewidth]{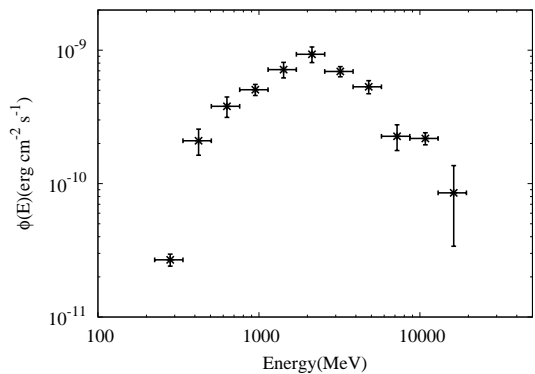}

\caption{Spectral energy distribution of {\gray}s  obtained for the NFW templates.  
}
\label{fig:nfw}
\end{figure*}

To compare our results with the  results reported by other groups, we  performed a likelihood fitting by using the NFW profile (see, e.g., \citet{daylan13}) and 2FGL catalog. The results are shown in Fig.\ref{fig:nfw}. The derived SED peaks at about 2 GeV,  and shows a very hard spectrum below the peak which is consistent with the spectra reported  e.g. by  \citet{daylan13} and \citet{macias14}. This consistency implies that the difference in the spectra  arising  from the bipolar  template,  is a real effect, but not a result of biases in the analysis.  

\begin{figure*}
\centering
\includegraphics[width=0.4\linewidth]{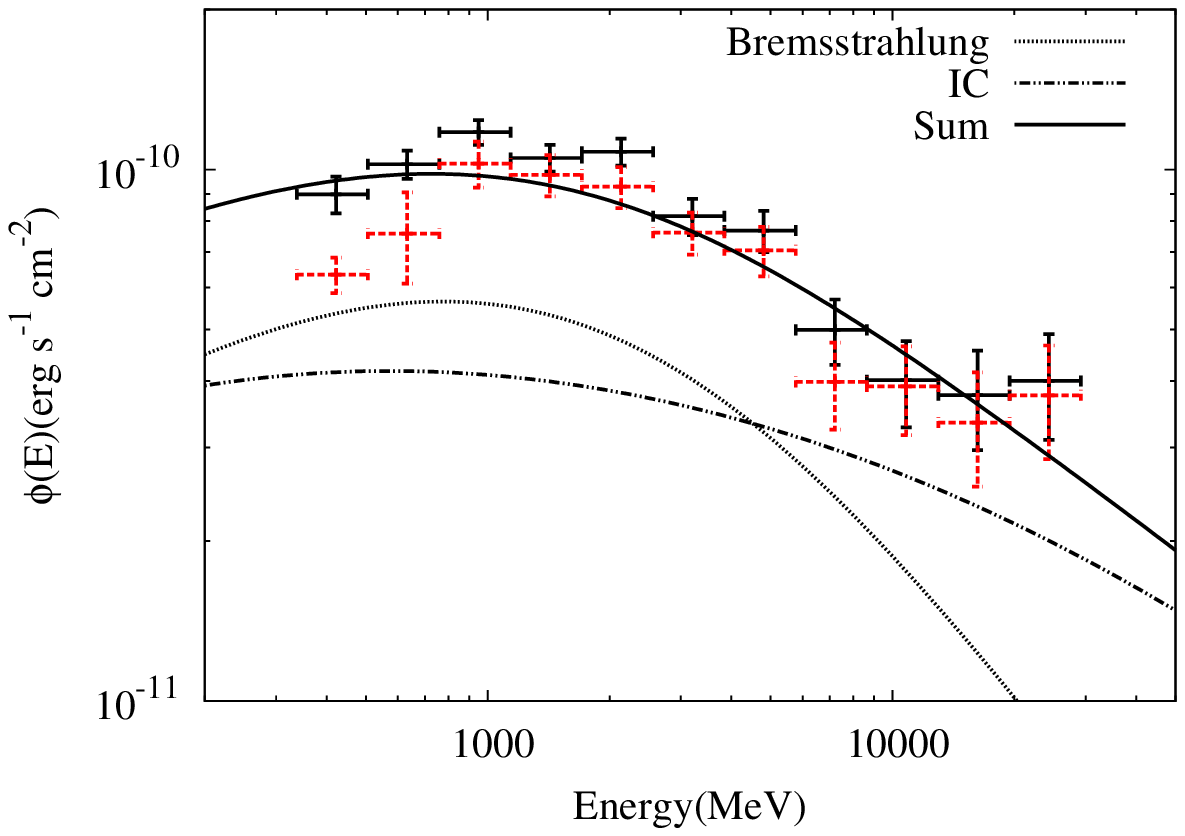}\includegraphics[width=0.4\linewidth]{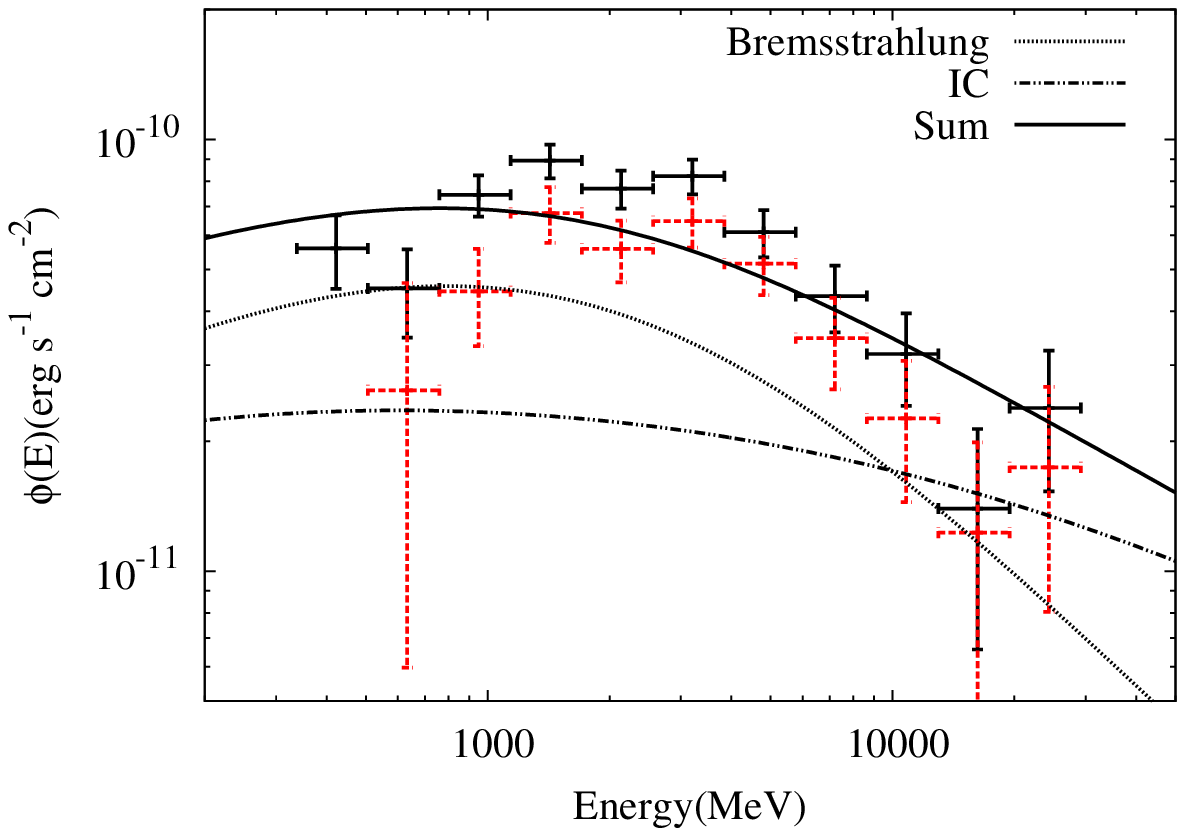}

\caption{Best fit  spectra of leptonic  $\gamma$-rays  for the northern  (left panel) and southern (right panel) parts of the excess. The red and black data points are the same as in Fig.2. The dotted  curves represent the  contribution from bremsstrahlung, the dot-dashed line represent ICs, and the solid curves are the summations.
 }
\label{fig:lep}
\end{figure*}

\begin{figure*}
\centering
\includegraphics[width=0.4\linewidth]{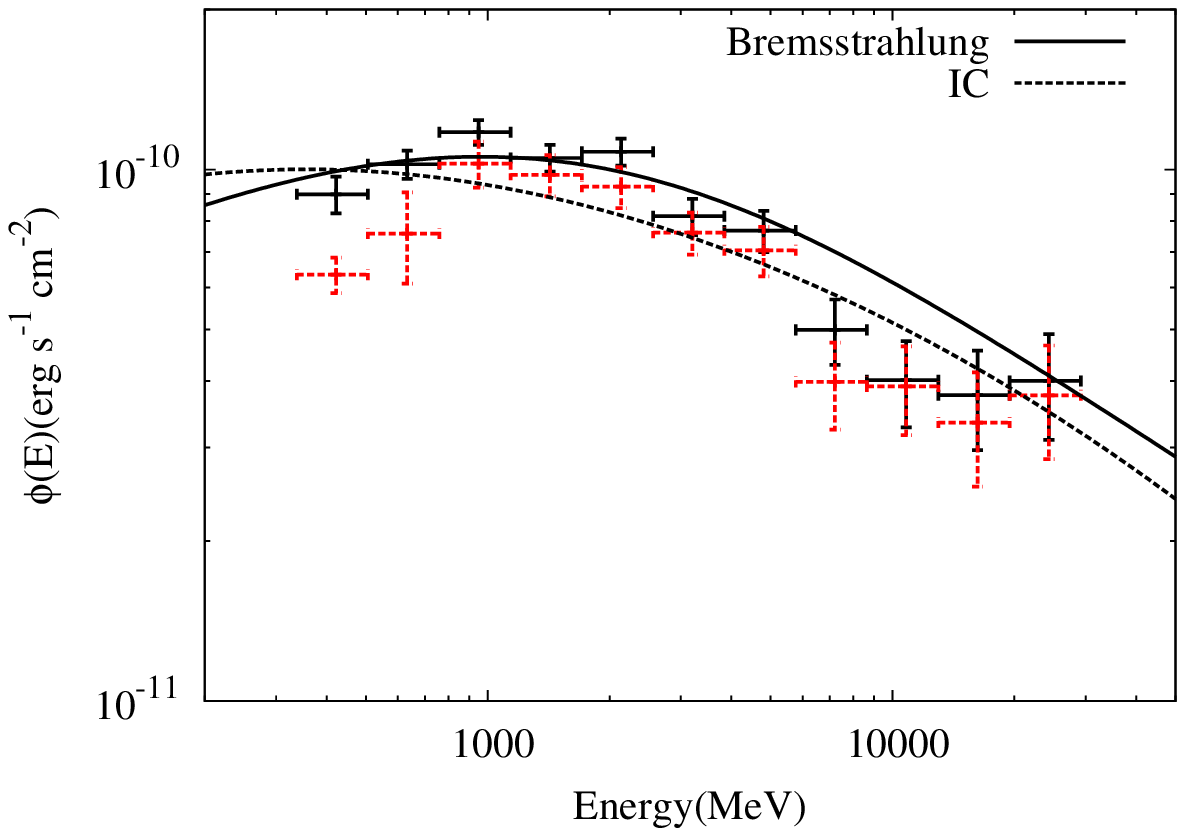}\includegraphics[width=0.4\linewidth]{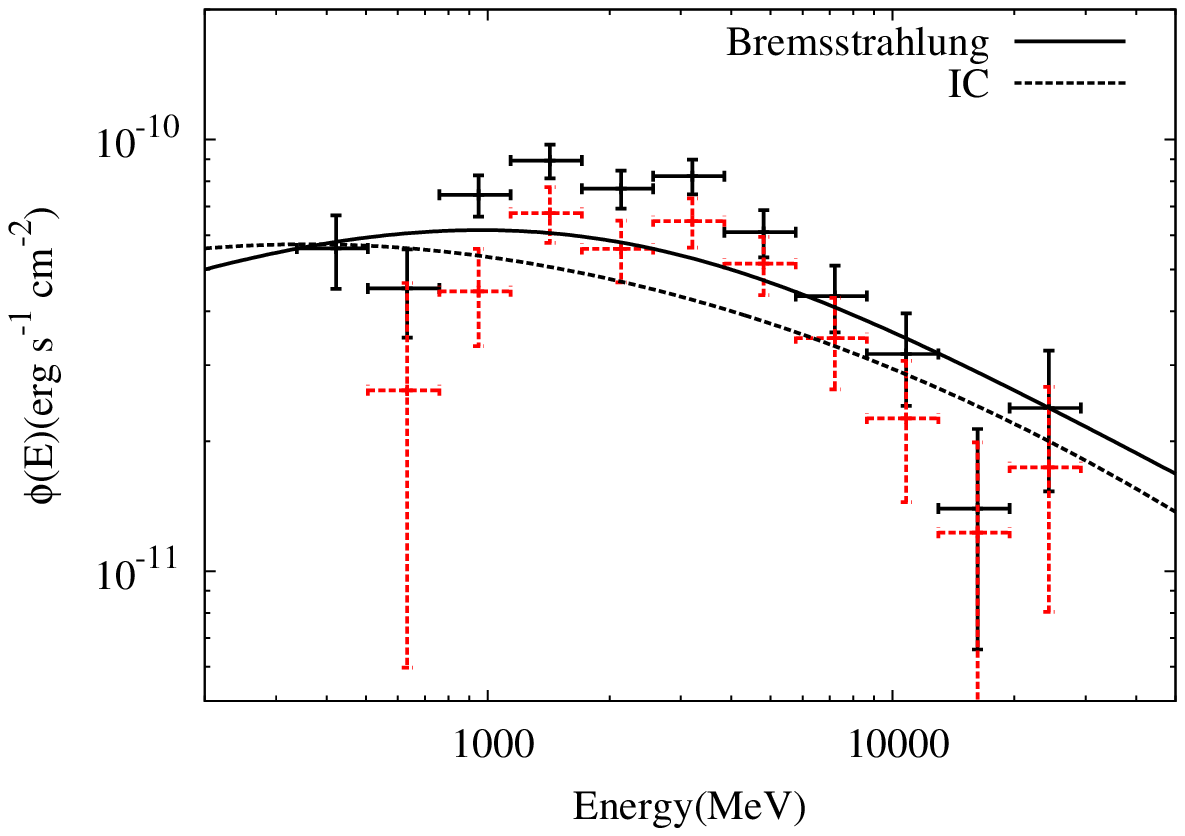}

\caption{Best fit spectra for the northern (left panel) and southern (right panel) parts of the excess by
interactions of electrons through the electron bremsstrahlung (solid curves) and the IC channels (dotted curves).
The red and black data points are the same as in Fig.2.  The curves for bremsstrahlung  and
ICs are normalized so as to explain the measurements separately(for details see the text).  %
}
\label{fig:icb}
\end{figure*}

\section{Radiation mechanism}\label{sec:discuss}
 Based on the claimed distinct spectral shape of the GeV  excess emission,  the  dark matter origin of 
 that emission  has been intensively discussed in recent years \citep{goodenough09, ak12, daylan13, macias14, weniger14,lacroix14} as an exciting,   although not a unique scenario \citep[e.g.,][]{carlson14, macias14_2, petrovic14,cholis15,abazajian15, huang15}. Our independent analysis 
of the  \fermi  data  confirms there is a new low  energy  $\gamma$-ray component, 
but not the previously reported spectral shape of radiation.   With  the spatial templates  
adopted in our study, we found  a smoother SED around 1 GeV that can be  readily explained by interactions of the cosmic ray protons and/or 
electrons without invoking specific or non-standard assumptions. Remarkably,   for  the given  radiation and gas densities, three radiation mechanisms 
could comparably contribute to the $\gamma$-ray production in this energy band:  (i) interactions of protons and  nuclei  with the ambient gas through the  production and decay of $\pi^0$-mesons,  (ii)  electron  bremsstrahlung and (iii)  IC scattering  of electrons.  
The average gas column density for the central 10-degree region  exceeds $10^{22} \ \rm cm^{-2}$  (Planck Collaboration  2011).  Assuming that it is  mostly contributed by the gas from the  central 1kpc region, the average hydrogen number density  in this region is estimated to be 
$\sim 3 \ \rm cm^{-3}$, which we then use as a fiducial value in further calculations of 
$\gamma$-ray production.  

(i) {\it $\pi^0$-decay gamma rays}

The  results shown in Fig.\ref{fig:SED},  indicate an apparent softening of the 
$\gamma$-ray spectra with an increase in energy  for both southern and northern 
parts of emission. This can be easily explained by assuming the proton spectrum in the following form
\begin{equation}
\psi(E_p)=N(E_p+E_{\rm 0})^{-\gamma}, 
\end{equation}
where $E_p$ is the proton kinetic energy.  We note that $E_{\rm 0}=1 ~\rm GeV$  implies a power-law spectrum in total energy: $E=E_p + m_pc^2$. The results of gamma-ray calculations for the northern and southern parts are shown in Fig.2. For the northern part, the best fit is  achieved  for  $E_{\rm 0} = 1~\rm GeV$ and $\gamma=2.4$, while for the southern  part  $E_{0} = 5~\rm GeV$ and $\gamma=2.7$.

The  spectral change in the proton  energy distribution  around $E_{\rm 0}$  in Eq.(1) can be for various reasons. In particular, it is possible that the CRs are accelerated in very dense regions where ionization losses,  which dominate at low energies,  make the spectrum harder \citep{deboer15}.  Secondly, the gas distribution may be very clumpy and the low-energy CRs cannot penetrate the dense cores of these clumps because of slower diffusion \citep{zirakashvili10,gabici14}.  Finally, such a spectral shape  with a cutoff (or break)   around  a few GeV can be  a result of a specific (e.g.) stochastic mechanism of  particle acceleration.  

(ii) {\it Leptonic origin of gamma rays}

Quite remarkably,  a break  at GeV energies is also required in the spectrum of electrons to explain the deficit in low-energy gamma rays within both  the
leptonic (bremsstrahlung and IC) channels.  Interestingly, such a low-energy cutoff, does exist  in the interstellar electron spectrum \citep{strong11, webber13} as well.  In particular, the electron spectrum derived from direct observations and from ratio data show an energy break at several GeV, with spectral indices 1.8 and 2.8 below and above $E_{\rm bk}$, respectively. These measurements can described by the following expression:
\begin{equation}
\psi(E_e)=N(E_e)^{-\gamma 1}(1+(\frac{E}{E_{bk}})^{\gamma 2-\gamma 1})^{-1} .\end{equation}
For the fit we leave $E_{bk}$ and $N$ as free parameters. 

Both bremsstrahlung  and IC scattering can significantly contribute to the \gray production. The emissivity of IC depends on the energy density of the interstellar radiation fields (ISRF). Here we adopt the value provided by GALPROP, in which the energy density of ISRF drops from $18.8 ~\rm eV/cm^3$  at a height $z=100$ pc to  $3.3 ~\rm eV/cm^3$ at  $z=1$ kpc. With such ISRF and an average gas density of about 3 $\rm cm^{-3}$,  the IC and bremsstrahlung channels give similar contributions to GeV {\gray}s.  Remarkably, 
the  electrons of approximately  same energy contribute to the  production of 100 MeV to 10 GeV  {\gray}s  through these two
channels. 

The results in Figure \ref{fig:lep}  show that the observed $\gamma$-ray fluxes can be explained by  a  combination  of  contributions  of  these two channels.  However,  the gas density  in this specific region contains non-negligible  uncertainties. It can be higher or lower than the one adopted in Figure \ref{fig:lep}. In that case the radiation would be dominated by the bremsstrahlung or IC channels, respectively.  In either case,  
with a slight adjustment  of the  parameters of the electron spectrum and the normalization factor in 
Eq.(1), we can explain  the  $\gamma$-ray  data. It is demonstrated  in Figure \ref{fig:icb}.

For both the  bremsstrahlung  and hadronic channels,  the $\gamma$-ray  emissivity is proportional to 
$n_{\rm CR} \times  n_{\rm gas} $, where $n_{\rm CR}$ and  $n_{\rm gas}$ are the CR (electrons or protons) and the gas densities, respectively. The recent measurements of the Planck satellite provide  the distribution of the  gas column density  throughout the Galaxy \citep{planck}. Thus,  using the information of the spatial distribution of the \gray flux,  we can derive the spatial distributions of the CR density.  For this reason, we divide the templates into slices with different latitudes ($|b|=[1^\circ,2^\circ],[2^\circ,3^\circ] ~\rm and~ [3^\circ, 5^\circ] $ for both northern and southern  parts ). For a better angular resolution we use data with energy of $\gamma$-rays exceeding  1 GeV. We  also exclude
the inner one-degree region from  this analysis to prevent  the contamination from the bright central source.  

\begin{figure*}
\centering
\includegraphics[width=0.4\linewidth]{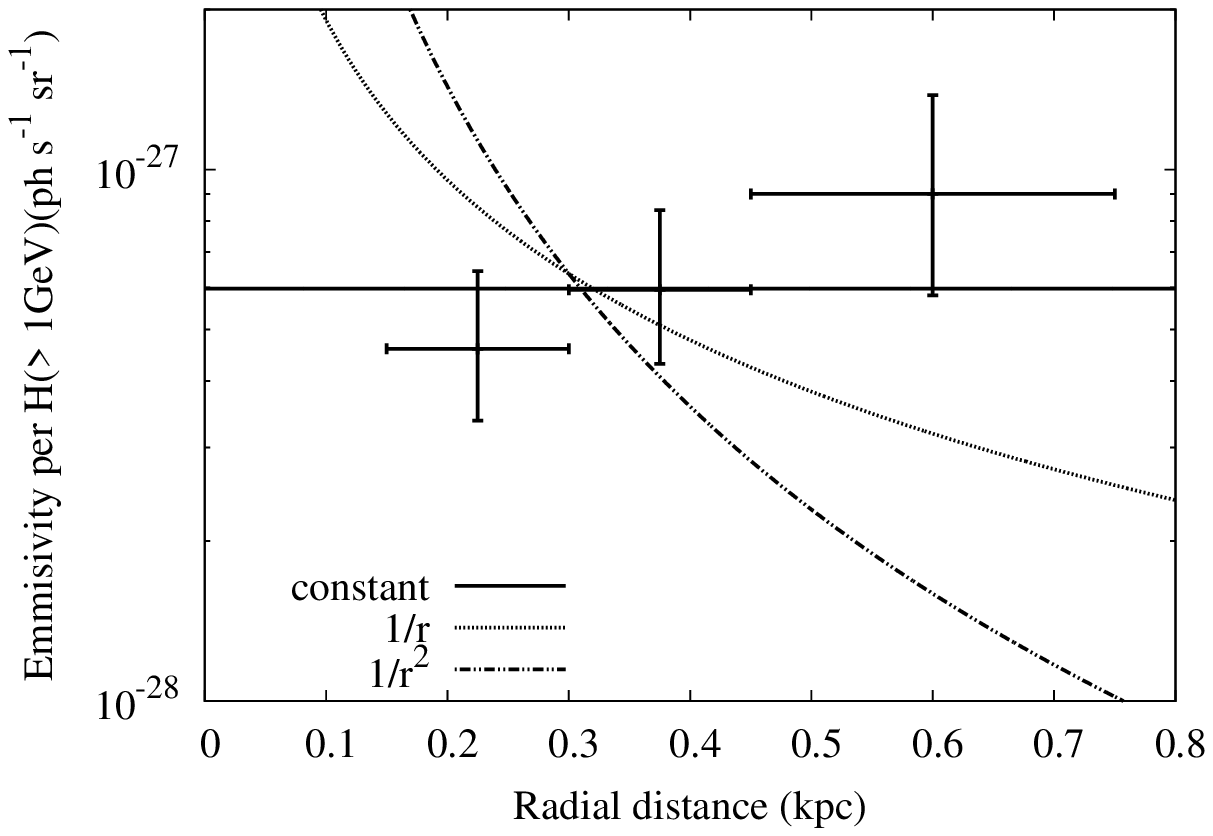}\includegraphics[width=0.4\linewidth]{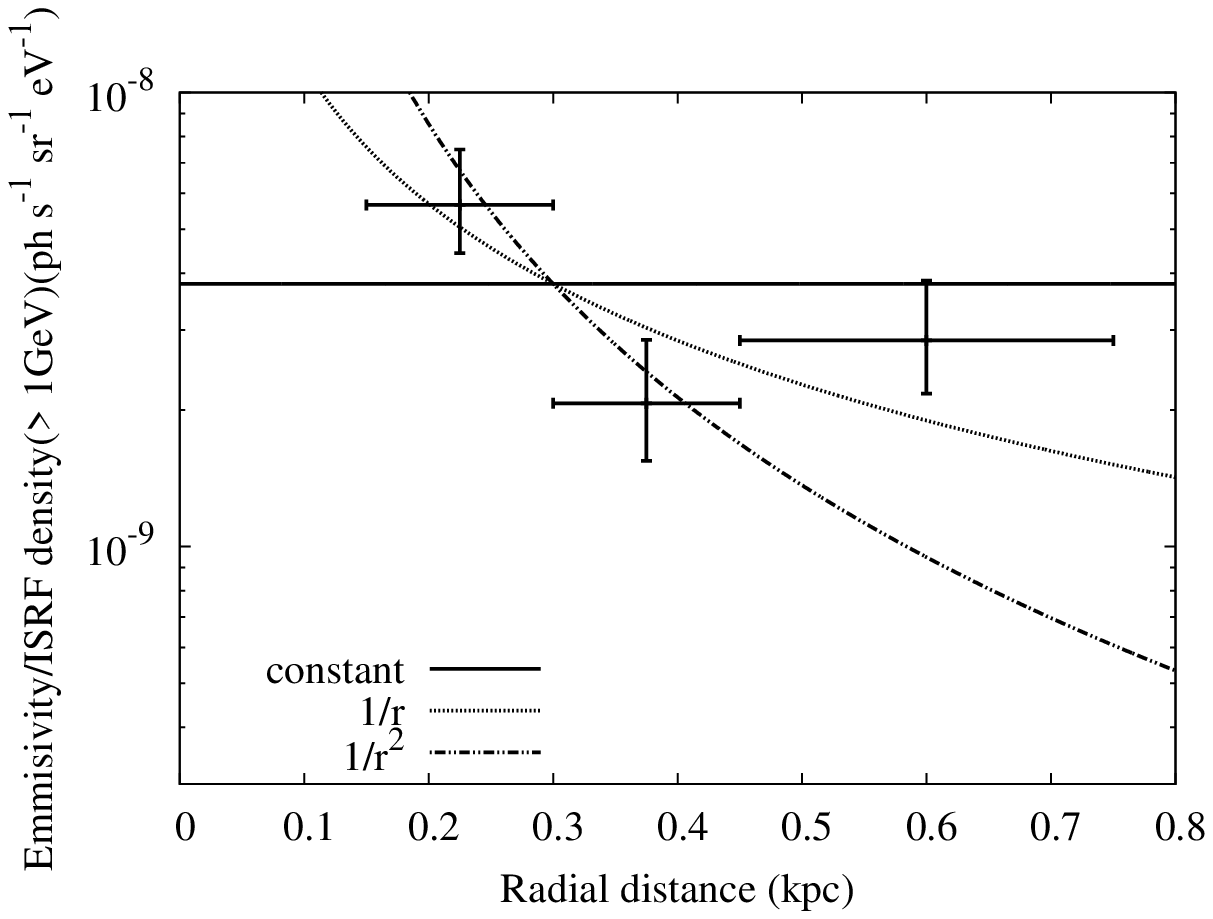}\

\caption{ Radial profiles of the \gray emissivity for the southern parts of the excess. For comparison,
we show the constant (solid curves), $1/r$ (dotted curves), and $1/r^2$ (dot-dashed curves) profiles,
which are predicted by the impulsive injection, continuous injection, and the transport through wind,
respectively. Left panel: the emissivity per hydrogen atom  related to interactions with gas (i.e. inelastic
interactions of protons and ions or  bremsstrahlung of electrons). Right panel: the \gray emissivity
due to IC normalized to the density of ISRF $1~\rm eV/cm^3$.
}
\label{fig:pro}
\end{figure*}

To derive the gas column density, we use the formula relating the dust opacity and the column density, using 
the  dust as the reference emissivity according to Eq.~(4) of the paper of \citet{planck}:
\begin{equation}\label{eq:dust}
\tau_M(\lambda) = \left(\frac{\tau_D(\lambda)}{N_H}\right)^{dust}[N_{H{\rm I}}+2X_{CO}W_{CO}],
 \end{equation}
where $\tau_M$ is the dust opacity as a function of the wavelength $\lambda$,  $(\tau_D/N_H)^{dust}$ is the reference dust emissivity measured in low-$N_H$ regions, $W_{CO}$ is the integrated brightness temperature of the CO emission, and $X_{CO}=N_{H_{2}}/W_{CO}$ is the so-called $H_2/CO$ conversion factor.
Substituting the latter into Eq.~(\ref{eq:dust}), one obtains
\begin{equation}
N_H = N_{H{\rm I}} +2 N_{H_2} =  \tau_m(\lambda)\left[\left(\frac{\tau_D(\lambda)}{N_H}\right)^{dust}\right]^{-1}. 
\end{equation}
We use a dust emissivity at $353~\rm GHz$ of $(\tau_D/N_H)^{dust}_{353{\rm~GHz}}=1.18\pm0.17\times10^{-26}$~cm$^2$,  from Table~3 of \citet{planck}. 
Then, to find the \gray emissivities per H-atom, we  divide the \gray flux by the column density.  This value should be proportional to the CR density.  The results for the southern part of the excess are summarized in Figure~\ref{fig:pro}. The results for northern part of the excess are similar. The spatial distribution of the CR density depends on the injection history. In the spherically symmetric case, assuming that the CR accelerator is located  in the GC region,  the impulsive injection predicts a constant distribution with the distance  $r$ , while the continuous injection predicts  a profile$~1/r$ . Finally, if CR transportation proceeds via wind, it should have a $~1/r^2$ profile. All the profiles are indicated in Fig. \ref{fig:pro}. One can see that the  radial distribution of the CR density derived from the spatial distributions of $\gamma$ rays and the gas is best explained  by an impulsive injection.  

In the IC scenario,  the  $\gamma$-ray emissivity is normalized to  $1 \rm eV/cm^3$ energy density of ISRF. To derive the radial distribution of this quantity, we  divided the gamma-ray flux by the ISRF density in each slice as described above. The results are shown in Figure \ref{fig:pro}. In this case,  a radial dependence  of the CR electron distribution closer to 1/r type profile is explained by the fact that  the distribution of  ISRF  is more homogeneous than the gas distribution.  We should mention the work of \citet{petrovic14} and \citet{cholis15}, who claim  that the excess of the radiation could  be mainly contributed by the IC scattering of electrons. However, the results in Figs. \ref{fig:SED}, \ref{fig:lep}, and  \ref{fig:icb} demonstrate that  at this stage we cannot give a preference to any of  three  alternatives related to  the $pp$, the  
electron bremsstrahlung and IC scenarios. 

Thus while the interpretation of the morphology of $\gamma$-rays within the $pp$ and the electron bremsstrahlung scenarios  demands a burst type injection,
the  interpretation of data within the IC scenario gives a preference to the continuous injection.   Apparently,  the question of the injection regime should be specified by the timescale of formation of these  large structures. If the particle propagation is dominated by diffusion, the characteristic timescale is  $t \sim R^2/6D \sim 10^6$ to $10^7$ yrs, depending on the  diffusion coefficient  $D$ of CRs  at GeV energies.  

Another important concern for  understanding the origin of these particles is the total energy budget.  The total luminosity of both the northern and southern  parts  is about $10^{37} ~\rm erg/s$. Then, the required CR energy budget for a hadronic scenario is  $10^{52} (3~ \rm cm^{-3}/n)~\rm erg$.  For the leptonic  scenarios, the energetics are  
somewhat (factor of 5 and 3)  less for the  electron bremsstrahlung and IC  channels, respectively. All these estimates exceed the typical CR energy release in a single supernova explosion by 1.5 to 2 orders of
magnitude. 

If the GeV excess is dominated by interactions of protons or electrons with the ambient gas,  so that an impulsive injection is preferred,  the required energy release in these low-energy  CRs can be related to the higher activity of the GC in the past, some  $10^{6-7}$ years ago.  This  agrees with the conclusion based on the recent studies of the H$\alpha$ emission in the so-called Magellanic stream \citep{bland-hawthorn13} .   On the other hand, the IC scenario  prefers a (quasi) constant injection,  so  the  GeV electrons  can be linked to other source populations, in particular to supernova remnants, stellar clusters, 
pulser wind nebulae, binary systems, {\it \emph{etc}}.

\section{Conclusions}\label{sec:conc}

We  performed a new analysis based on the  5.5-years of observations of \fermi in the $10^{\circ} \times 10^{\circ}$ region towards the GC with a  focus on the so-called GeV excess reported previously.  We found  that the morphology of this extra component of {\gray}s  would 
prefer  a bipolar to a spherically symmetric structure.  Moreover, the $\gamma$-ray energy spectra appear smoother than reported in the previous studies. We argued that the energy spectra can be fitted by both hadronic and leptonic channels of \gray production.  In the case of leptonic origin of this radiation component, the contributions from interactions of electrons with the ambient gas and radiation fields  are comparable. But the dominance of  either the electron bremsstrahlung or the IC scattering cannot be excluded given the uncertainties of the gas density in the production region.  The total energy budget in a parent's  charged particles in the  leptonic scenarios is less, by a factor of  a few,  than in  protons.  The  estimates  exceed   the  CR energy that can be provided by a single SNR  by 1.5
to 2 orders of magnitudes.   

Alternatively,   these particles can be  related to other source populations, in particular to pulsars (or pulsar wind nebulae), to the relativistic jets of binary systems, to powerful stellar winds in compact stellar clusters, and of course, to  Sgr A*, the central supermassive black hole  in the GC. Both the hadronic and the electron bremsstrahlung scenarios lean towards a burst type injection of  particles (over the last $10^{6-7}$ years),  which would favour a possible link of these particles to a high activity of Sgr A*,  On the other hand, the IC scenario, which favours a continuous injection,  is not excluded.   To distinguish between these possibilities we need a comprehensive modelling of the propagation and radiation of GeV protons and electrons in the central several kpc region of the inner Galaxy. Such studies can  provide important  insight into the origin of the GeV excess  and its relation to other CR  accelerators in that most active part of our Galaxy.

\bibliographystyle{aa}
\bibliography{ms_lan}
%\bibliography{27550}
\end{document}